\documentclass{IEEEtran}

\usepackage{cmap}
\usepackage[T1]{fontenc}
\usepackage[utf8]{inputenc}
\ifdefined\pdfgentounicode \input{glyphtounicode} \pdfgentounicode=1 \fi

\usepackage{cite}
\usepackage{amsmath,amssymb,amsfonts}
\usepackage{amsthm}
\usepackage{booktabs}
\usepackage{multirow}
\usepackage{array}
\usepackage{tabularx}
\usepackage{graphicx}
\usepackage{xcolor}
\usepackage{microtype}
\usepackage{url}
\usepackage{algorithm}
\usepackage{algpseudocode}
\usepackage{placeins}
\usepackage{hyperref}

\usepackage{ifmtarg}
\makeatletter
\renewcommand{\Call}[2]{\textproc{#1}\@ifmtarg{#2}{}{(#2)}}
\makeatother

\newcolumntype{Y}{>{\raggedright\arraybackslash}X}
\newcolumntype{R}{>{\raggedleft\arraybackslash}X}

\newtheorem{invariant}{Invariant}

\hypersetup{
  colorlinks=true,
  linkcolor=blue!60!black,
  citecolor=blue!60!black,
  urlcolor=blue!60!black,
  pdftitle={Feasibility of Homomorphic Inference for a Genomic Foundation Model},
  pdfauthor={Christos Galanopoulos; Kimon Antonios Provatas; Ilias Georgakopoulos-Soares},
  pdfsubject={Client-assisted homomorphic inference for genomic sequence analysis},
  pdfkeywords={genomic privacy; homomorphic encryption; privacy-preserving inference; transformers}
}
\graphicspath{{../figures/}{figures/}}

\begin{document}

%

\title{Feasibility of Homomorphic Inference for a Genomic Foundation Model}

\author{Christos Galanopoulos, Kimon Antonios Provatas, and Ilias Georgakopoulos-Soares%
\thanks{C. Galanopoulos is with The University of Texas at Austin, Austin, TX, USA.}%
\thanks{K. A. Provatas and I. Georgakopoulos-Soares are with The University of Texas at Austin
and The University of Texas at Austin College of Pharmacy, Austin, TX, USA.}%
\thanks{Corresponding author: I. Georgakopoulos-Soares (e-mail: ilias@austin.utexas.edu).}%
}

\maketitle

\begin{abstract}
Human genomic sequences can identify individuals, cannot be replaced after disclosure, and are the
inputs that genomic foundation models are designed to interpret. We assess whether a compute
provider can execute a released genomic foundation model without receiving query-derived genomic
values in plaintext and whether
correctness, memory, or cost prevents complete encrypted inference. We first reproduce the released
model on three genomic task families and freeze an independently validated numerical reference. We
then implement a client-assisted approximate homomorphic encryption protocol: the provider evaluates
linear algebra on ciphertexts, while the key-holding data owner evaluates exact normalization,
causal softmax, and activation functions at fixed boundaries. A non-interactive configuration
completes one released-weight block but exceeds the tested accelerator-memory envelope when
configured for composition. The client-assisted configuration executes all released transformer
blocks and the task head for one held-out genomic-signal input at its full prompt length. It matches
the frozen final label, peaks at 9{,}839 mebibytes of accelerator memory, and completes in 6{,}683
seconds on one accelerator. These results establish arithmetic feasibility for a complete
classifier, while repeatability, network transport, and private token-index lookup remain
unresolved. The biomedical significance is that, under the stated threat model, a served genomic
model can process an encoded sequence without exposing plaintext query-derived activations to the
compute provider.
\end{abstract}


%

\section{Introduction}

Human genomic sequence carries a persistent confidentiality risk. It can identify its contributor
and reveal information about relatives; re-identification has been demonstrated after conventional
identifiers were removed~\cite{gymrek2013,erlich2018}. A disclosed genome cannot be replaced like a
password. European data-protection law accordingly treats genetic data as a special
category~\cite{gdpr}. Yet genomic foundation models derive their value from this same sequence:
DNAGPT supports classification, regression, and generation through a common transformer
backbone~\cite{dnagpt2023}. Homomorphic encryption offers access to a served model without sending
the provider plaintext genomic values.

DNAGPT's public release makes this experiment auditable and permits local execution. We use it as a
reproducible surrogate for a genomic model offered only as a service by policy or contract, accessed
by a CPU-equipped data owner without accelerator hardware. We ask whether its released computation
can execute faithfully within the cryptographic and memory budget, and whether the resource cost
supports a target use. These are separate feasibility and practicality verdicts: a correct but slow
encrypted path can resolve the former while leaving the latter open.

The technical difficulty is concentrated in transformer nonlinearities. CKKS supports approximate
packed arithmetic for dense projections, rotations, and accumulations~\cite{ckks2017}, whereas
non-interactive LayerNorm, softmax, and GELU require polynomial approximations that increase depth
and resident cryptographic material. Moving exact functions to the key holder resets the depth
budget but introduces interaction and a different threat model. THOR demonstrates non-interactive
homomorphic transformer inference; Safhire uses the server-linear/client-nonlinear division in
convolutional networks~\cite{thor,safhire2025}. We study the correctness, memory, and executed work
of the latter established pattern on a released genomic transformer.

We first evaluate genomic-signal recognition, promoter and splice-site classification, and mRNA
abundance regression, freezing outputs before encrypted work. An independent NumPy reference is
asserted against upstream PyTorch at $2\times10^{-5}$ for every released block and the task head.
We then build non-interactive and client-assisted CKKS designs, with fixed, non-adaptive boundaries
for exact nonlinearities in the latter. Instrumented evaluations record correctness and memory;
runtime assertions preserve the cryptographic operation schedule during systems optimization.

The client-assisted protocol executes all $12$ released blocks and the task head for one held-out
input at the $103$-token genomic-signal prompt. Its label matches the frozen reference, with head
margin relative error $8.56\times10^{-9}$. The non-interactive design completed a released-weight
block but exceeded the tested $80$~GB envelope while loading its composition context and evaluation
keys. The complete client-assisted run instead peaks at $9{,}839$~MiB of process GPU memory and
takes $6{,}683$~s on one accelerator. Bounded caching also prevents encoded plaintexts from
accumulating across stages. These results establish arithmetic feasibility at the stated encrypted
boundary while leaving deployment practicality a separate question.

The paper makes the following contributions:

\begin{itemize}
  \item An independently verified plaintext reference across three genomic task families.
  \item A specified client-assisted CKKS protocol, including packing, depth resets, algorithms,
        and a semi-honest threat model.
  \item Complete released-weight inference for one held-out input, with label and intermediate
        verification under a predeclared tolerance.
  \item Fixed-circuit systems optimization with an asserted operation schedule, bounded host
        caching, and separate feasibility and practicality verdicts.
\end{itemize}

\begin{figure*}[!t]\centering
  \includegraphics[width=\textwidth]{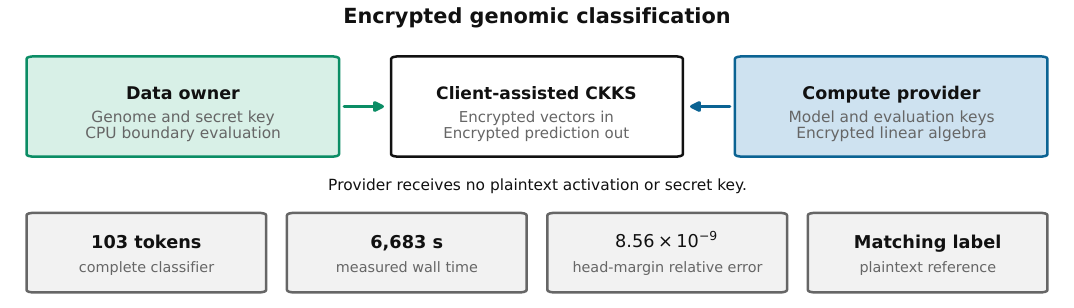}
  \caption{Study overview and measured complete-classifier result. The data owner retains the
  secret key and evaluates exact nonlinearities; the compute provider evaluates encrypted linear
  algebra. The reported execution begins at encrypted embedded vectors and covers one genomic-signal input.}
  \label{fig:overview}
\end{figure*}


\section{Background}

\subsection{DNAGPT and the genomic task families}
DNAGPT is an autoregressive transformer for sequence and numerical tasks over DNA~\cite{dnagpt2023}.
The evaluated $0.1$-billion-parameter configuration has $12$ blocks, hidden width $768$, $12$
causal-attention heads of width $64$, and feed-forward width $3{,}072$. Its input combines
$6$-mer DNA tokens with task-format tokens.

Each block applies LayerNorm; query, key, and value projections; causal self-attention; an output
projection and residual; a second LayerNorm; and an expanded MLP with GELU and another residual.
Dense maps use fixed weights; LayerNorm, attention, and GELU introduce data-dependent products or
nonlinear functions.

Released heads cover genomic-signal recognition and mRNA abundance regression; the
genome-understanding benchmark uses a locally fine-tuned linear head (Section~\ref{sec:baseline}).
These tasks follow convolutional signal recognition~\cite{deepgsr}, abundance
regression~\cite{xpresso}, and the encoder-based benchmark~\cite{dnabert2}. They establish model
fidelity before encrypted evaluation of the signal-recognition path.

\subsection{CKKS}
CKKS supports approximate arithmetic on packed real or complex vectors~\cite{ckks2017}. Values
occupy plaintext slots encrypted under the data owner's key. Addition and multiplication act
slotwise; rotations move values between slots. This single-instruction, multiple-data model
expresses public matrix transforms through rotated ciphertexts, plaintext diagonals, and accumulation.

Query--key scoring and attention context instead multiply encrypted query-derived operands,
requiring ciphertext--ciphertext multiplication and relinearization. Public weights and masks
remain plaintext; query-derived operands remain encrypted at the server.

Rescaling controls fixed-point scale using levels from a finite modulus chain. Bootstrapping
refreshes the budget without decryption; fresh client encryption resets it interactively. Their
system costs make multiplicative depth an architectural resource.

\subsection{Why the nonlinearities are the architecture}
LayerNorm requires inverse square root, softmax requires exponentiation and normalization, and the
MLP requires GELU. A non-interactive circuit approximates these functions polynomially over declared
input ranges. Higher degree consumes additional depth, increasing the context and evaluation-key
material needed across repeated blocks. Exact evaluation at declared client boundaries removes
range calibration and resets depth, while requiring an online client that observes intermediate
activations. This choice determines memory, interaction, and the threat model;
Section~\ref{sec:protocol} specifies both designs.

\subsection{Software backend selection}
OpenFHE provides CKKS context and security-parameter construction, encoding, rotations, polynomial
evaluation, and approximate bootstrapping~\cite{openfhe}, establishing reference behavior for local
operators and the non-interactive design. The evaluated GPU backend interoperates with those
representations and supplies every measured server-side CUDA operation. Python, NumPy, and PyTorch
construct and validate the plaintext reference outside the encrypted performance path.

We also considered Concrete ML and the Boolean/integer-oriented TFHE-rs
ecosystem~\cite{concreteml,tfhers}. Concrete ML documents a hybrid language-model protocol with
client-side attention and activations and server-side linear layers~\cite{concreteml-inference}.
This was a documentation-based assessment, not a local DNAGPT performance comparison. We retained
the validated CKKS implementation; this engineering choice does not establish CKKS as the only
viable encryption scheme.


\section{Related work}

\subsection{Private transformer inference}

Gazelle combines homomorphic linear layers with garbled-circuit
nonlinearities~\cite{gazelle2018}. Iron, BOLT, Nimbus, and BumbleBee adapt homomorphic encryption,
secret sharing, and/or oblivious-transfer-based protocols to transformer attention,
normalization, and activations~\cite{iron2022,bolt2024,nimbus2024,bumblebee2025}. They optimize
joint secure computation; our server evaluates the linear path in CKKS and the data owner
evaluates exact nonlinearities at explicit boundaries, changing both the security contract and
the allocation of work.

The Safhire preprint describes a closely related architecture: its client decrypts an intermediate,
evaluates the nonlinearity, and returns fresh encryption~\cite{safhire2025}. It studies convolutional
networks and adds randomized output permutation intended to hinder reconstruction from intermediates.
We apply this established pattern to a released genomic transformer at its task-defined prompt,
preserving released weights and exact nonlinearities. The contribution is the independently
checked verification chain and measured arithmetic, boundary schedule, and memory behavior.

THOR demonstrates full non-interactive homomorphic transformer evaluation using diagonal-major
organization and compact packing~\cite{thor}. Thus, our observed memory wall is specific to the
evaluated backend, parameters, packing, and circuit. Client assistance tests whether released
computation can be preserved when the data owner is available for exact nonlinear evaluation.

\subsection{Homomorphic encryption for genomics}

The iDASH competition evaluated encrypted association studies, sequence search, and related
genomic workflows~\cite{idash2018}. Private genomic-query systems combine homomorphic encryption,
hashing, and set intersection to test for variants without disclosing the genome or
query~\cite{privategenomicqueries2017}. Beyond these searches and prescribed statistics, DNAGPT
inference composes dense projections, attention, normalization, and feed-forward layers while
preserving the numerical behavior of a released foundation model.

\subsection{GPU-accelerated homomorphic encryption}

The evaluated GPU backend supplies number-theoretic transforms, key switching, rotations, and
ciphertext arithmetic interoperable with the host runtime~\cite{gpu-ckks-backend,openfhe}.
Its missing ciphertext serialization prevented process-sharded evaluation, leaving thread-level
concurrency and in-process device placement. Multi-GPU transformer work motivates further
placement and communication--computation overlap~\cite{aegis}; adopting those schedules requires
direct evaluation of serialization and transport, which this implementation does not exercise.

\subsection{Homomorphic matrix primitives}

Matrix layout determines rotations, plaintext encodings, and reductions. THOR specializes it for
transformers~\cite{thor}; HE-BLAS supplies homomorphic matrix--vector, vector--vector, and
matrix--matrix reductions implemented with BLAS~\cite{heblas2025}. These motivate less repeated
packing, fewer diagonal copies, and shape-matched reductions (Section~\ref{sec:optimization}).
Because packing, keys, levels, and the operation schedule are coupled, candidate primitives must
be revalidated against the frozen reference and schedule before performance comparison.


\section{Study Design and Threat Model}
\label{sec:threat-model}

\subsection{Deployment roles and privacy objective}
Figure~\ref{fig:overview} summarizes the deployment and complete-classifier evaluation.
The deployment has two parties but one cryptographic confidentiality objective. The data owner
holds a human genomic sequence, ordinary CPU resources, and no GPU. It will not disclose the
sequence to the compute provider. The model owner operates the accelerator infrastructure and
serves the DNAGPT model rather than distributing it to clients. Keeping the weights server-side is
an operational preference of this deployment, not a model-confidentiality guarantee supplied by
the protocol.

The released DNAGPT artifact makes the experiment reproducible; it does not itself require this
service arrangement. A client may run the public release locally. We use it as an auditable
surrogate for a genomic model whose owner limits access to a service by policy or contract, and the
deployment conclusions apply under that premise.

The protected asset is therefore the genome. The data owner encrypts its embedded numeric vectors
and retains the secret key; the compute provider evaluates the served model on those ciphertexts.
The protocol is relevant only where these roles are fixed. If a client is allowed to receive and
run the model, local plaintext inference is the simpler architecture.

\subsection{Local inference as a comparison}
Where model distribution is permitted, the client can avoid all cryptographic work by running the
ordinary forward pass. Client assistance cannot provide a computational advantage over that
option. Its purpose is to make a service-only model available without sending the genomic input in
plaintext.

That comparison changes the deployment premise. A served model is available to the data owner only
through the provider's interface; handing the client the weights creates a different service. Under
the premise studied here, the available choices are to disclose the genome, decline the inference,
or evaluate through a privacy-preserving protocol. Client-assisted CKKS addresses the last choice.
The reason to use it is access to a served model without genomic disclosure, not an attempt to beat
plaintext inference.

\begin{table*}[t]\centering\small
\caption{Resource and trust allocation under local and served-model deployment.}
\label{tab:deployments}
\begin{tabularx}{\textwidth}{Y c c}
\toprule
 & Client-local inference & Client-assisted CKKS \\
\midrule
Client must hold the model & yes & no \\
Server observes the genome & not applicable & no \\
Client requires a GPU & deployment-dependent & no \\
Server evaluates model linear algebra & no & on ciphertexts \\
Client evaluates model nonlinearities & as part of local model & at declared boundaries \\
\bottomrule
\end{tabularx}
\end{table*}

The protocol assigns encrypted dense linear algebra to the party with accelerator hardware while
the key holder evaluates exact nonlinearities on CPU. This allocation protects the genomic input
without requiring the data owner to operate a GPU.

This allocation becomes more server-heavy as model width grows. At fixed sequence structure, the
client's boundary work scales with the number of activation values, whereas the server's dense
transforms scale quadratically with hidden width. The client fraction therefore decreases as
$O(1/D)$ with hidden width $D$ under this fixed sequence structure. This is an asymptotic property
of the work allocation, not a measurement of a wider DNAGPT model.

\subsection{Adversary model}
We consider a semi-honest compute provider under $128$-bit classical CKKS parameters. It follows
the prescribed circuit but may inspect everything it legitimately receives: public and evaluation
keys, model parameters, ciphertexts, sequence length, message sizes, operation order, boundary
schedule, and timing. It does not alter ciphertexts to probe the client or deviate from the
declared boundary sequence.

Under this adversary, the compute provider receives no plaintext genomic vector, plaintext
activation, partial decryption, or secret-key material. At a boundary it supplies a ciphertext and
receives a fresh ciphertext under the data owner's public key. The guarantee concerns the
provider's view of query-derived values; public model structure and plaintext weights remain
available to the provider as inputs to ciphertext--plaintext operations.

\begin{invariant}[Key custody]
The data owner generates and retains the secret key. No protocol message contains the key or a
partial decryption, and the compute provider has no decryption operation.
\end{invariant}

\begin{invariant}[No plaintext at the server]
Every query-derived value held by the compute provider is a ciphertext under the data owner's key.
Client-evaluated nonlinearities return through fresh encryption rather than as decoded values.
\end{invariant}

\begin{invariant}[Non-adaptive boundaries]
The circuit and sequence length determine every boundary before execution; decrypted content does
not select the next operation. The implementation asserts the complete schedule and boundary counts
for each accepted run and aborts on deviation. Transcript equality across multiple private inputs
of equal length has not been measured as a separate leakage experiment.
\end{invariant}

\subsection{What is not claimed}
Model confidentiality is not claimed. The data owner observes full intermediate activations at
every nonlinearity. Those observations decouple the network into shallow segments with known
nonlinearities, turning global model inversion into inexpensive per-segment regression. Keeping
weights on the server may deter casual copying, but it does not prevent extraction by a determined
client. This concession does not change the genomic-data guarantee, which protects the data owner
from the compute provider.

Malicious servers, authenticated transport, production key custody, traffic analysis, timing and
physical side channels, and compromised clients are outside the evaluated boundary. Encrypted
scope begins after token-index lookup, so private lookup is also unresolved. The data owner sees
its own intermediate activations and final output; privacy from the key holder is not an objective
of this protocol.


\subsection{Plaintext reference and experimental design}
\label{sec:baseline}

An encrypted execution is useful only if the computation it reproduces is itself meaningful and
independently checked. Task metrics answer the first question: they establish that the released
model and locally trained heads retain the genomic capabilities for which the model is being
evaluated. Per-example predictions answer the second. They are recorded before any encrypted work
and become the frozen plaintext reference against which encrypted outputs are accepted.

The plaintext harness imports the upstream model implementation and preserves its classification
and regression forward paths. It adds batching, metric computation, and result recording without
changing model arithmetic. Evaluation uses the canonical test split for each task, and each run
records both its configuration and the per-example outputs needed downstream. Agreement with a
published aggregate metric is evidence that the intended model behavior survived local
reproduction; the recorded output for the selected encrypted input is the stricter numerical
contract used by the cryptographic evaluation.

The reference is not trusted merely because it is written in NumPy. Released checkpoint weights
are evaluated through the upstream PyTorch block and, independently, through a float64 NumPy
re-derivation. For every one of the $12$ transformer blocks, the two outputs are asserted equal at
relative and absolute tolerance $2\times10^{-5}$; the classifier head is checked by the same
procedure. Generation aborts on any divergence. Only after this upstream-to-reference comparison
passes are the weights and reference output supplied to the encrypted circuit. This creates two
independent checks: the plaintext re-derivation must reproduce the released model, and the
encrypted computation must reproduce that validated re-derivation.

\subsection{Task-derived sequence lengths}
The encrypted anchor length follows from the genomic-signal task rather than from a convenient
cryptographic setting. Its $600$-base-pair window becomes $100$ sequence tokens under the model's
$6$-mer tokenization; three task-format special tokens bring the complete prompt to $103$ tokens.
The evaluated prompt is therefore the task input consumed by the released classifier, rather than
a shortened surrogate chosen to fit the encrypted implementation.

The same arithmetic determines the other classification prompts. Core-promoter detection maps a
$70$-base-pair window to $12$ sequence tokens plus one special token, or $13$ total. The
$300$-base-pair promoter task uses $50$ sequence tokens plus one special, or $51$; the
$400$-base-pair splice task uses $67$ sequence tokens plus one special, or $68$. The
$103$-token genomic-signal prompt is the longest of these encrypted-scope tasks and consequently
provides the anchor for the packing and sequence-length analysis in Sections~\ref{sec:protocol}
and~\ref{sec:results}.

\section{Client-assisted CKKS protocol}
\label{sec:protocol}

\begin{figure*}[!t]\centering
  \includegraphics[width=\textwidth]{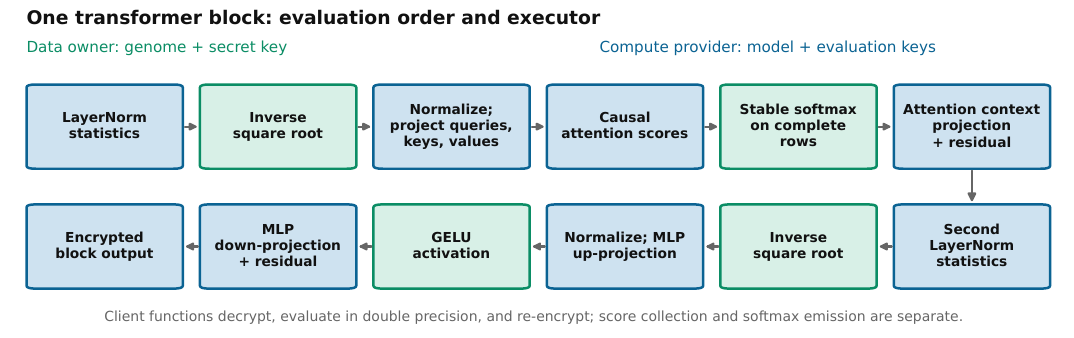}
  \caption{Client-assisted block evaluation. The compute provider evaluates encrypted linear algebra
  and the data owner evaluates exact nonlinearities at declared boundaries. The provider has public
  weights, keys, and execution metadata, but receives no plaintext query-derived activation or secret key.}
  \label{fig:architecture}
\end{figure*}

Table~\ref{tab:notation} defines the evaluated configuration. Figure~\ref{fig:architecture} locates
the server computation and the client calls within a transformer block. The protocol retains the
released weights and exact nonlinearities while bounding encrypted depth through fresh client
encryption.

\begin{table}[!t]\centering\small
\caption{Notation. All values are those of the evaluated configuration.}
\label{tab:notation}
\begin{tabularx}{\columnwidth}{l l Y}
\toprule
Symbol & Value & Meaning \\
\midrule
$D$        & 768    & hidden width of the released model \\
$H$        & 12     & attention heads, of $D/H = 64$ dimensions each \\
$D_{\text{mlp}}$ & 3072 & MLP expansion width, $4D$ \\
$T$        & 103    & tokenized prompt length for the genomic-signal task \\
$B$        & 8      & tokens packed per ciphertext (token lanes) \\
$G$        & 13     & ciphertext groups, $\lceil T/B \rceil$ \\
$W$        & 1024   & feature width padded to a power of two \\
$C$        & 4      & activation copies per ciphertext, $D_{\text{mlp}}/D$ \\
$N$        & 65536  & polynomial ring degree \\
$N/2$      & 32768  & usable plaintext slots per ciphertext, $C \cdot W \cdot B$ \\
$L$        & 13     & multiplicative depth \\
$\lambda$  & 128    & classical security level, in bits \\
\bottomrule
\end{tabularx}
\end{table}

\subsection{Client boundaries and normalization}
CKKS carries affine maps, reductions, residual additions, and attention products. The inverse
square root in LayerNorm, causal softmax, and GELU use declared client calls. At each call, the
data owner decrypts values derived from its query, evaluates the function in double precision, and
encrypts the result afresh under the same key. Algorithm~\ref{alg:boundary} gives this operation.

\begin{algorithm}[H]
\small
\caption{The client boundary. The data owner is the only party holding $\mathsf{sk}$, and every
value it decrypts derives from its own query.}
\label{alg:boundary}
\begin{algorithmic}[1]
\Function{ClientBoundary}{$c$, $f$}
  \State $v \gets \Call{Decode}{\Call{Decrypt}{\mathsf{sk},c}}$
  \State $v' \gets f(v)$ \Comment{evaluated exactly, in double precision}
  \State \Return $\Call{Encrypt}{\mathsf{pk},\Call{Encode}{v'}}$ \Comment{fresh ciphertext at
         level $0$}
\EndFunction
\end{algorithmic}
\end{algorithm}

Fresh encryption returns the activation to level~$0$. The configured depth of $13$ therefore bounds
the longest encrypted run between boundaries rather than the depth accumulated across all $12$
transformer blocks. The ciphertext lineage remains at the server between these calls.

For LayerNorm, the server computes the mean and centered variance under encryption and adds
$\varepsilon$ before sending the variance term to the client. Only the inverse square root is
evaluated there; the server then applies the returned factor and public scale, as specified in
Algorithm~\ref{alg:layernorm}. No bias is omitted as an approximation: the released checkpoint
contains no bias tensors.

\begin{algorithm}[H]
\small
\caption{LayerNorm. The server computes both statistics under encryption; only the inverse square
root crosses the boundary. The released checkpoint carries no bias term.}
\label{alg:layernorm}
\begin{algorithmic}[1]
\Function{LayerNorm}{$x$, $\gamma$}
  \State $\mu \gets \Call{ReduceSum}{x}\odot(1/D)$
  \State $z \gets x - \mu$
  \State $\sigma^2 \gets \Call{ReduceSum}{z \odot z}\odot(1/D) + \varepsilon$
  \State $r \gets \Call{ClientBoundary}{\sigma^2,\ t \mapsto t^{-1/2}}$
  \State \Return $(z \odot r) \odot \gamma$
\EndFunction
\end{algorithmic}
\end{algorithm}

\subsection{Packing and dense transforms}
\begin{figure}[!t]\centering
  \includegraphics[width=\columnwidth]{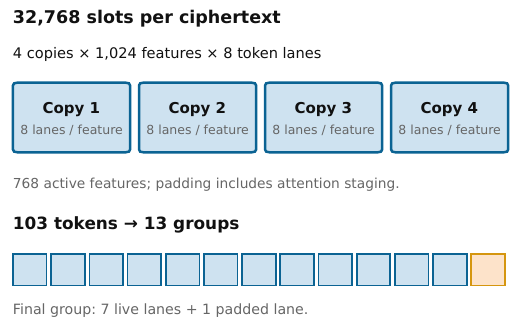}
  \caption{Slot layout and grouping for the measured prompt. The final ciphertext group carries
  seven live tokens and one padded lane; the groups produce the lower-triangular causal score tiles.}
  \label{fig:packing}
\end{figure}

The slot layout in Fig.~\ref{fig:packing} assigns $32{,}768$ slots as $4$ activation copies by
$1{,}024$ padded features by $8$ token lanes. Token lane is the innermost index: whole-lane
rotations move features without mixing tokens, and smaller rotations align token lanes. The
released width of $768$ occupies the active feature region; padding supports the power-of-two
transform and attention-score staging. Rows and columns outside the active width contribute zeros.

The four copies express the MLP expansion from width $768$ to $3{,}072$ as parallel chunks.
The $103$ tokens occupy $13$ ciphertext groups. A $32\times32$ baby-step giant-step decomposition
covers the $1{,}024$ padded features. Algorithm~\ref{alg:matmul} combines shared baby rotations
with encoded weight diagonals, cloning each plaintext for use at the receiving ciphertext's level.

\begin{algorithm}[H]
\small
\caption{Packed matrix product, baby-step giant-step over the packed diagonals.
$N_1 = N_2 = 32$ and $N_1N_2 = W = 1024$.}
\label{alg:matmul}
\begin{algorithmic}[1]
\Function{MatMul}{$\beta$, $M$}
  \State $d \gets \Call{EncodedDiagonals}{M,\ \mathrm{level}(\beta_0)}$
         \Comment{encoded once, cloned per use}
  \State $y \gets 0$
  \For{$j \gets 0$ \textbf{to} $N_2-1$}
    \State $u \gets \sum_{i<N_1} \beta_i \odot \Call{Clone}{d_{N_1 j + i}}$
    \State $y \gets y + (j = 0\ ?\ u : \Call{Rot}{u, B N_1 j})$
  \EndFor
  \State \Return $y$
\EndFunction
\end{algorithmic}
\end{algorithm}

Eight-token packing reduces the serial-equivalent dense-product count from $1{,}236$ to $156$,
a structural reduction of $7.92\times$. Rotations, masking, encoding, and client work remain, so
this operation-count ratio does not specify a latency speedup.

\subsection{Block evaluation}
\label{sec:algorithms}
Algorithm~\ref{alg:block} follows the released block through normalization, query/key/value
projection, causal attention, output projection, residuals, and the expanded MLP. The cached
diagonals are flushed after the last use of each weight set; this changes storage lifetime without
changing the arithmetic. Baby rotations are shared among transforms that consume the same input.

\begin{algorithm}[!t]
\small
\caption{Evaluating one transformer block. $G=13$ token groups; $\odot$ is homomorphic
multiplication; every operation outside a \textsc{Client} call runs on ciphertexts.}
\label{alg:block}
\begin{algorithmic}[1]
\Require encrypted activations $c_0,\dots,c_{G-1}$; public weights $W$
\Statex \textit{Stage 1: normalization and projection}
\For{$g \gets 0$ \textbf{to} $G-1$}
  \State $n \gets \Call{LayerNorm}{c_g,\ \gamma_1}$ \Comment{Algorithm~\ref{alg:layernorm}}
  \State $\beta \gets \Call{BabyRotations}{n}$ \Comment{$N_1-1$ rotations, shared}
  \State $Q_g \gets \Call{MatMul}{\beta,W_q}$; $K_g \gets \Call{MatMul}{\beta,W_k}$
  \State $V_g \gets \Call{MatMul}{\beta,W_v}$
\EndFor
\State \Call{FlushEncodedWeights}{} \Comment{bounds host memory to one stage}
\Statex \textit{Stage 2: causal attention scores}
\For{$k \gets 0$ \textbf{to} $G-1$}
  \State $\tilde{K} \gets \{\Call{LaneShift}{K_k,\delta}\}$ for each $\delta$ required by a
         later query group
  \For{$q \gets k$ \textbf{to} $G-1$}
    \State $S \gets 0$
    \For{$\delta \in \Call{ActiveShifts}{q,k}$}
      \State $P \gets Q_q \odot \tilde{K}_\delta$
      \For{$h \gets 0$ \textbf{to} $H-1$}
        \State $S \gets S + \Call{ReduceSum}{P \odot \mathrm{mask}_h} \odot
               \mathrm{isolate}(q,k,\delta,h)/\sqrt{d_h}$
      \EndFor
    \EndFor
    \State \Call{Client.TakeScoreTile}{$S,q,k$} \Comment{decrypt only}
  \EndFor
\EndFor
\State \Call{Client.Softmax}{} \Comment{stable softmax over complete causal rows}
\Statex \textit{Stage 3: attention context}
\For{$k \gets 0$ \textbf{to} $G-1$}
  \State $\tilde{V} \gets \{\Call{LaneShift}{V_k,\delta}\}$
  \For{$q \gets k$ \textbf{to} $G-1$; $\ \delta \in \Call{ActiveShifts}{q,k}$}
    \State $A \gets \Call{Client.EmitWeightTile}{q,k,\delta}$ \Comment{re-encrypt only}
    \State $C_q \gets C_q + A \odot \tilde{V}_\delta$
  \EndFor
\EndFor
\Statex \textit{Stages 4 and 5: projection, MLP, residuals}
\For{$g \gets 0$ \textbf{to} $G-1$}
  \State $R_g \gets c_g + \Call{MatMul}{\Call{BabyRotations}{C_g},W_{\text{proj}}}$
\EndFor
\State \Call{FlushEncodedWeights}{}
\For{$g \gets 0$ \textbf{to} $G-1$}
  \State $\beta \gets \Call{BabyRotations}{\Call{LayerNorm}{R_g,\gamma_2}}$
  \State $h \gets \sum_{c<4} \Call{MatMul}{\beta,W_{\text{fc}}^{c}} \odot \mathrm{copy}_c$
  \State $a \gets \Call{Client.GELU}{h}$
  \State $c'_g \gets R_g + \sum_{c<4}
         \Call{MatMul}{\Call{BabyRotations}{a \odot \mathrm{copy}_c},W_{\text{mlp}}^{c}}$
\EndFor
\State \Return $c'_0,\dots,c'_{G-1}$
\end{algorithmic}
\end{algorithm}

The protocol transcript is fixed by the circuit and sequence length. Each block makes $857$
physical client boundary crossings, including $91$ score-tile decryptions and $727$ weight-tile
encryptions. They carry $129{,}162$ logical nonlinear instances. Softmax accounts for $128{,}544$,
or $99.5\%$, of them, concentrating the boundary work in attention. The implementation asserts
the operation schedule, boundary counts, and remaining depth at run time and aborts on deviation.

\subsection{Causal attention}
Attention visits the $91$ lower-triangular pairs of the $13$ token groups. Shifts required by
later query groups are computed once per key group and reused. The client collects all encrypted
score tiles before stable softmax normalizes complete causal rows, then returns fresh encrypted
weight tiles for context aggregation. This normalization is a global synchronization point.

Both query--key scoring and weight--value aggregation multiply two activation-derived operands.
Of the block's $1{,}506$ ciphertext--ciphertext multiplications, LayerNorm contributes $26$
centered squares and $26$ products with returned inverse square roots. Attention contributes
$727$ query--key and $727$ weight--value products, or $97\%$ of the total.

\subsection{Depth, parameters, and composition}
The retained context has ring dimension $65{,}536$, $32{,}768$ usable slots, and a $128$-bit
classical security level. It uses $50$-bit scaling, a $60$-bit first modulus, $3$ hybrid
key-switching digits, and $80$ rotation keys. Depth $13$ is the smallest demonstrated passing
depth for these parameters and packing; the block finishes with $6$ levels remaining.

Between blocks, the data owner decrypts the full hidden state, restores its token-vector layout,
and encrypts fresh level-$0$ inputs under the same context and key lineage. The preliminary
two-block execution at two tokens reached relative error $8.48\times10^{-11}$ without GPU-memory
growth in the second block. Section~\ref{sec:results} reports the complete task-length composition.

\subsection{Non-interactive comparison}
The non-interactive design replaces every nonlinearity with a polynomial approximation and
retains a single ciphertext lineage without intermediate decryption. It completes one
released-weight block. When configured for composition, however, its context and evaluation keys
exceed the tested $80$~GB GPU-memory envelope before useful composition arithmetic begins.

Repeated polynomial evaluation requires a deeper modulus budget and a bootstrapping schedule
whose associated material must remain resident. This memory limit applies to the tested library,
parameters, packing, and hardware. The client-assisted design shortens encrypted segments by
evaluating the exact functions at the client and refreshing through fresh encryption.

%

\section{Systems Optimization at Fixed Circuit}
\label{sec:optimization}

The encrypted circuit remains fixed throughout this section: $156$ dense products,
$177{,}734$ ciphertext--plaintext multiplications, $1{,}506$ ciphertext--ciphertext multiplications,
$8{,}173$ rotations, and $857$ client boundary crossings per block. Runtime assertions check these
counts, the boundary schedule, and remaining depth before the MLP, aborting on deviation. Packing,
depth, reference, and acceptance criterion are unchanged. Optimizations alter host-data preparation,
residency, and CPU placement without removing arithmetic or moving linear work to the client.
The retained system completes the full classifier in one clean sample. No pre-optimization timing
was measured under matching conditions, so we report this endpoint directly rather than as a
speedup factor.

\subsection{Host-side encoding redundancy}
The baseline rebuilt and encoded $1{,}024$ diagonals for each of $156$ dense products, yielding
$159{,}744$ encodings per block. About $90\%$ were byte-identical because the $12$ weight matrices
recur across $13$ token groups. Each diagonal must be packed, scaled, and encoded at the receiving
ciphertext's level before GPU multiplication can begin. Caching this model-invariant representation
therefore removes redundant host preparation while preserving encrypted products and accumulations.

\subsection{Encode once, clone per use}
The retained cache encodes each distinct weight diagonal into a pristine host template indexed by
matrix and multiplicative level. Each multiplication receives a clone sharing encoded host data
by reference but carrying fresh device state, loaded and evicted for that operation. This removes
about $90\%$ of the $159{,}744$ encodings without persistent GPU plaintext objects.

Cloning follows the backend's object semantics: its load path returns early for an already-loaded
object, so direct reuse across levels can expose stale residue-number-system limbs. Fresh device
state forces loading at the current level. An isolated GPU test verified clone-versus-fresh-encode
equality, cross-level reuse, stable memory during repeated multiplication, and parallel-versus-serial
encoding equality before application to a complete block.

\subsection{Bounding host memory}
An early cache retained templates across completed stages and failed allocation when the MLP
constructed its templates above those left by projection and attention.

Two exact changes bound residency. Templates are encoded at their level of use, retaining only
required limbs; the backend would otherwise drop a level-$0$ plaintext to that same level.
The cache is then flushed after query/key/value projection and after attention projection, where
the respective weights reach their final use. The group-major loop retains reuse within each stage;
eviction between stages cannot create additional re-encoding.

\begin{figure}[t]
\centering
\includegraphics[width=\columnwidth]{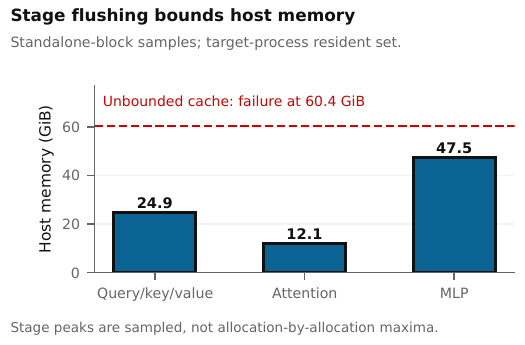}
\caption{Host-memory bounding through level-specific encoding and stage-local template lifetimes.
The measured resident set falls after projection templates are flushed; the MLP determines the
retained complete-block peak.}
\label{fig:memory}
\end{figure}

The resident-set trace (Fig.~\ref{fig:memory}) reaches $24.9$~GiB during query/key/value projection
and $12.1$~GiB during attention after projection templates are flushed. The MLP's eight matrices
determine the $47.5$~GiB block peak. Thus, the largest live stage determines peak host memory.

\subsection{Placement and parallelism}
The retained implementation binds threads to physical cores and uses NUMA-local allocation for
encoding and GPU staging buffers. Decryption, nonlinear evaluation, and re-encryption share those
host resources. Template construction uses dynamic scheduling across $32$ cores: independent
diagonals are encoded in parallel without changing encrypted multiplication order. Output equality
was checked against serial construction. The individual timing effects of affinity and batching
await sourced paired measurements.

\begin{table*}[t]\centering\small
\caption{The optimization campaign. Every row leaves the operation schedule, packing,
multiplicative depth, and frozen plaintext reference unchanged.}
\label{tab:optimization}
\begin{tabularx}{\textwidth}{Y l l}
\toprule
Change & Verified effect & Status \\
\midrule
Re-encode every plaintext per multiplication & $159{,}744$ encodings & baseline \\
Thread affinity, NUMA-local allocation & host placement changed & retained \\
Encode once, clone per use & removes about $90\%$ of encodings & retained \\
Parallel batched encoding & equal to serial encoding & retained \\
Encode at use level, flush between stages & peak $47.5$~GiB RSS & enabling \\
\midrule
\textbf{Net} & \textbf{checked operation schedule unchanged} & \\
\bottomrule
\end{tabularx}
\end{table*}

\subsection{Remaining optimization targets}
About $17{,}400$ mask plaintexts are still encoded afresh per block. Head, score-isolation, lane,
and copy-activity masks are fixed by the layout; caching them remains unimplemented.

The four packed activation copies currently carry duplicates through dense transforms. An analyzed
layout applies distinct transforms to them, combining query/key/value projections and MLP expansion
chunks. At fixed model arithmetic, the projected schedule reduces dense products from $156$ to
$52$ and dense ciphertext--plaintext multiplications from $159{,}744$ to $53{,}248$, lowering the
total ciphertext--plaintext count by roughly $60\%$ before mask and copy-repair overhead. An exact
reconstruction proof is required before implementation.

Softmax accounts for $99.5\%$ of boundary values, with independent cryptographic work per tile.
Concurrent boundaries were unsafe under the shared crypto context, requiring sequential execution;
a thread-safe context would permit concurrency.


\section{Results}
\label{sec:results}

We evaluate the client-assisted circuit along four axes: plaintext model fidelity, encrypted
numerical agreement, executed work, and resource cost. The strongest result is one complete
classifier execution for a held-out genomic-signal input at the task's full prompt length.

\subsection{Plaintext model fidelity}

\begin{table*}[!t]\centering\small
\caption{Released DNAGPT evaluated locally on three genomic task families. Published comparators
are DNAGPT-M for signal recognition and abundance regression, and DNABERT-2 for the three
understanding-benchmark tasks. Signal recognition and abundance regression use released fine-tuned
heads; the understanding-benchmark rows use a locally fine-tuned head, as no head was released.}
\label{tab:baseline}
\begin{tabularx}{\textwidth}{Y l l l l}
\toprule
Task & Metric & This work & Reference & Source \\
\midrule
Polyadenylation-signal recognition$^{\ddagger}$ & accuracy & 0.9124 & 0.9151 & \cite{dnagpt2023} \\
 & F1 & 0.916 & --- & \\
Core promoter detection & MCC & 0.680 & 0.69 & \cite{dnabert2} \\
300\,bp promoter detection & MCC & 0.897 & 0.87 & \cite{dnabert2} \\
Splice-site classification & MCC & 0.831 & 0.85 & \cite{dnabert2} \\
mRNA abundance regression$^{\dagger}$ & $r^2$ & 0.562 & 0.62 & \cite{dnagpt2023} \\
 & Pearson $r$ & 0.753 & --- & \\
\bottomrule
\end{tabularx}
\vspace{2pt}
{\footnotesize $^{\ddagger}$The reference is DNAGPT's own reported accuracy for this model class,
measured on a held-out quarter of the set; this work evaluates all $22{,}604$ examples. The
comparison is same-model but not same-split.\\
$^{\dagger}$Model-fidelity evidence only. This task is outside the encrypted
scope; see Section~\ref{sec:limitations}.}
\end{table*}

\begin{figure}[!t]\centering
  \includegraphics[width=\columnwidth]{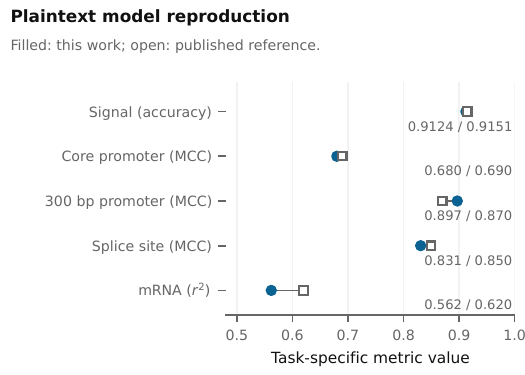}
  \caption{Plaintext model fidelity against published references. The signal-recognition comparison
  uses different test splits; promoter and splice-site comparisons use the stated benchmark splits.
  Abundance regression is outside the encrypted scope.}
  \label{fig:baseline}
\end{figure}

Table~\ref{tab:baseline} and Fig.~\ref{fig:baseline} show that the target carries useful signal across classification and
regression tasks. On the complete polyadenylation-signal test set of $22{,}604$ examples, the
released classifier reaches $0.9124$ accuracy and $0.916$ F1, against the $0.9151$ accuracy
reported for this model class on a held-out quarter of that set~\cite{dnagpt2023}. The locally
fine-tuned understanding-benchmark heads reach MCC $0.680$, $0.897$, and $0.831$ on core-promoter,
promoter, and splice-site detection, against the $0.69$, $0.87$, and $0.85$ reported on the same
splits~\cite{dnabert2}. The released abundance-regression head reaches $r^2=0.562$ and Pearson
$r=0.753$, against the $0.62$ result reported by DNAGPT on the Xpresso dataset (Xpresso's own
published human $r^2$ is $0.59$~\cite{xpresso})~\cite{dnagpt2023}. These results establish model fidelity; encrypted evaluation is still
required to establish cryptographic feasibility.

Two dataset qualifications belong with these baseline results. DNAGPT does not release a head for
the understanding benchmark, so we trained a linear head and evaluated the three human promoter
and splice-site datasets selected for this study. For abundance regression, the historical source
was recovered and its split procedure reproduced, but the resulting test set may not be
gene-for-gene identical to the historical split. The abundance result is therefore model-fidelity
evidence and is not used as an encrypted target.

\subsection{Complete-classifier correctness}

Correctness rests on the two-stage chain in Section~\ref{sec:baseline}. Before encryption, the
independent NumPy reference is asserted against the upstream PyTorch computation at tolerance
$2\times10^{-5}$ for every transformer block and the classifier head. The encrypted circuit is
then compared with that validated reference under the predeclared relative infinity-norm tolerance
of $4\times10^{-2}$.

All $12$ released transformer blocks, $11$ full-hidden-state client refreshes, and the released
task head complete for the $103$-token input. Across the block outputs, global relative error is at
most $2.07\times10^{-8}$, and worst-token relative error is at most
$1.54\times10^{-7}$. The final encrypted-head margin has relative error
$8.56\times10^{-9}$ and yields the same label as the frozen plaintext reference. Every block output
stays within the predeclared tolerance after its refresh. This is an encrypted prediction for one
held-out input, not an encrypted estimate of task accuracy over the test set.

The complete result subsumes the earlier composition checks summarized in
Table~\ref{tab:scope}. A separate two-block execution at two tokens had established that a
full-hidden-state refresh starts the next block at level~$0$ without increasing GPU memory. The
task-length execution now verifies that mechanism across every released block and the task head.

\begin{figure}[!t]\centering
  \includegraphics[width=\columnwidth]{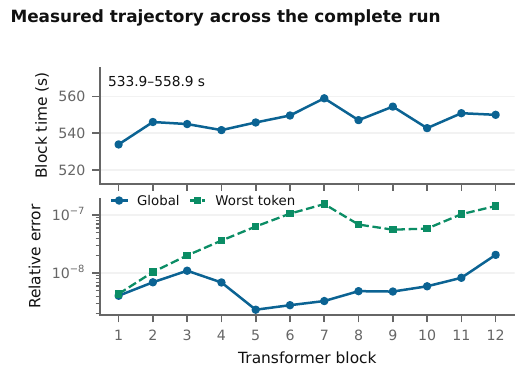}
  \caption{Blockwise time and numerical agreement in the measured complete-classifier execution.
  The consecutive blocks belong to one input and one run; their variation is not independent-run variance.}
  \label{fig:timeline}
\end{figure}

Figure~\ref{fig:timeline} shows the blockwise timing and numerical agreement within this execution.

\begin{table*}[!t]\centering\small
\caption{Encrypted evaluation at the reported scopes. The complete-classifier row represents one
held-out genomic-signal input, not a test-set accuracy measurement.}
\label{tab:scope}
\begin{tabularx}{\textwidth}{>{\hsize=1.25\hsize}Y >{\hsize=.85\hsize}Y >{\hsize=1.2\hsize}Y >{\hsize=.5\hsize}Y >{\hsize=1.2\hsize}Y}
\toprule
Scope & Status & Numerical agreement & Wall time & Process peak memory \\
\midrule
Two blocks, 2 tokens, client refresh & measured & $8.48\times10^{-11}$ after block 2 & --- & no growth in block 2 \\
One block, 103 tokens & measured & $4.64\times10^{-9}$ & --- & $9.6$~GiB GPU \\
Twelve blocks and task head, 103 tokens & measured & label match; $8.56\times10^{-9}$ head margin & $6{,}683$~s & $9{,}839$~MiB GPU; $49.7$~GiB host \\
Encrypted token-index lookup & outside evaluated boundary & --- & --- & --- \\
\bottomrule
\end{tabularx}
\end{table*}

\subsection{Executed work}

Table~\ref{tab:cost} reports the checked operation schedule. Every block retains the same schedule
used throughout the systems campaign. The complete run adds the released task head and asserts its
own totals, so the final label cannot be attributed to silently omitting a model stage.

\begin{table*}[!t]\centering\small
\caption{Checked work for one task-length block and for the measured complete classifier. Physical
client crossings are in-process cryptographic operations, not network messages.}
\label{tab:cost}
\begin{tabularx}{\textwidth}{Y r r l}
\toprule
Operation & One block & Complete classifier & Executor \\
\midrule
Dense encrypted products & 156 & 1{,}872 blocks $+$ 1 head & compute provider \\
Ciphertext--plaintext multiplications & 177{,}734 & 2{,}133{,}839 & compute provider \\
Ciphertext--ciphertext multiplications & 1{,}506 & 18{,}076 & compute provider \\
Rotations & 8{,}173 & --- & compute provider \\
Slot reductions & 8{,}776 & --- & compute provider \\
Physical client boundary crossings & 857 & 10{,}299 & data owner \\
Logical nonlinear values & 129{,}162 & 1{,}551{,}081 & data owner \\
\bottomrule
\end{tabularx}
\end{table*}

For one block, the $857$ physical boundary crossings carry $129{,}162$ logical values. Softmax
accounts for $128{,}544$, or $99.5\%$, of them; LayerNorm and GELU account for the remainder.
Eight-token packing reduces the serial-equivalent dense-product count from $1{,}236$ to $156$, a
$7.92\times$ structural reduction. Rotations, masking, client work, and host preparation remain,
so that factor is an operation-count reduction rather than a latency speedup.

\subsection{Measured elapsed time and work split}

The complete classifier takes $6{,}683$~s ($1.86$~h) from process start through completion;
encrypted evaluation accounts for $6{,}594.75$~s. The $12$ blocks consume $6{,}565.99$~s, the
encrypted head $9.05$~s, and the refreshes $19.71$~s. The block times within this execution range
from $533.9$ to $558.9$~s, with a mean of $547.2$~s. This within-run range is about
$4.5\%$ of the maximum; it describes block consistency in one execution, not independent-run
variance.

Server-side encrypted linear algebra accounts for $5{,}638.56$~s, or $85.5\%$ of encrypted
evaluation. The in-process client boundaries account for $956.19$~s, or $14.5\%$, and require no
GPU. The measured execution is therefore server-dominated, while the data owner's work remains
CPU-only. The client fraction should decrease as $O(1/D)$ with hidden width $D$ at fixed sequence
structure because boundary values scale with $D$ and server dense transforms with $D^2$; this is a
scaling argument, not a measurement of a wider model.

\begin{figure}[!t]\centering
  \includegraphics[width=\columnwidth]{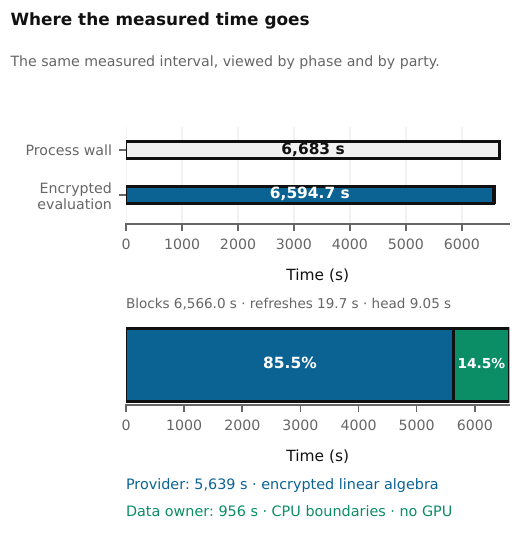}
  \caption{Measured complete-run time, decomposed by phase and by party. Top: process wall against
  instrumented transformer, refresh, and task-head subtotals; the remaining interval falls outside
  those subtotals. Bottom: the same encrypted-evaluation interval split between the compute
  provider's encrypted linear algebra and the data owner's in-process client boundaries, which run
  on CPU and exclude network transport.}
  \label{fig:cost-decomposition}
\end{figure}

Figure~\ref{fig:cost-decomposition} decomposes this same measured interval two ways: by phase
(process wall against blocks, refreshes, and the task head) and by responsible party (server
against client).

\subsection{Memory}

The complete process peaks at $9{,}839$~MiB of GPU memory and $49.7$~GiB of host resident memory.
The GPU value is the evaluating process footprint, not device-wide occupancy. At one-block scope,
host memory follows the encoded-template lifetime: query/key/value projection reaches $24.9$~GiB,
attention begins after a cache flush at $12.1$~GiB, and the MLP reaches $47.5$~GiB while holding the
largest live weight set.

The non-interactive composition configuration and the host cache expose different memory
mechanisms. The former exceeded its tested $80$~GB GPU envelope while loading a deeper context and
evaluation keys. Client re-encryption bounds multiplicative depth between declared boundaries and
thereby moves that accelerator barrier. Separately, stage-bounded eviction prevents encoded host
templates from accumulating; an unbounded cache was terminated at $60.4$~GiB before the retained
policy reduced the block peak to $47.5$~GiB.

\subsection{Sequence-length boundary}

Genomic-signal recognition uses $600$ base pairs, which become $100$ sequence tokens under
$6$-mer tokenization; three task-format tokens produce the $103$-token measured prompt. Eight-token
packing maps it to $13$ ciphertext groups. Core-promoter, promoter, and splice-site prompts use
fewer groups under the same layout, but their complete encrypted classifiers have not been
measured; any elapsed-time estimate for them is a fixed-circuit projection.

The abundance-regression task lies outside the encrypted scope. Its $10{,}500$ base pairs produce
$1{,}755$ tokens and $220$ ciphertext groups. The causal schedule would require $24{,}310$ score
tiles, placing it in a different circuit-size regime from the classification prompts.

\begin{figure}[!t]\centering
  \includegraphics[width=\columnwidth]{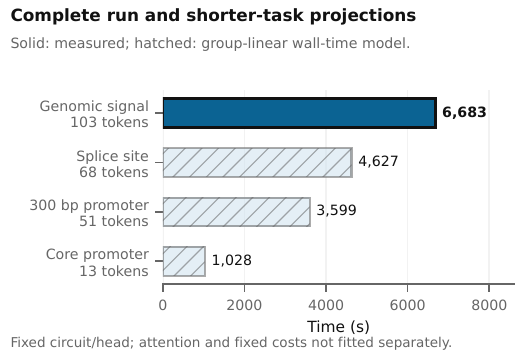}
  \caption{Sequence-length comparison. The genomic-signal classifier is the measured complete run;
  hatched values assume whole-program time proportional to ciphertext-group count. This illustrative
  model holds the circuit and head fixed and does not separately model quadratic attention or fixed overhead.}
  \label{fig:scaling}
\end{figure}

Figure~\ref{fig:scaling} scales the measured complete-program time by ciphertext-group count for
an illustrative comparison. The alternative task classifiers and their heads have not been measured;
the projections do not separately model quadratic attention cost or fixed overhead.

\subsection{Feasibility}
\label{sec:verdict-feasibility}

Arithmetic feasibility is established for one complete released-weight genomic-signal classifier
at its full prompt length. Every transformer block, every declared refresh, and the released head
complete within the tested memory envelope; the final label matches the frozen plaintext reference;
and intermediate errors remain far inside the acceptance tolerance. This verdict applies from
encrypted embedded vectors through the released head; it does not cover private token-index lookup
or prediction agreement over the full genomic-signal test set.

\subsection{Practicality, judged separately}
\label{sec:verdict-practicality}

The measured cost is substantial: one encrypted example takes $1.86$~h on one A100-class GPU with
CPU-only client assistance. That value is a real latency measurement for the evaluated program,
not a projection. It is also one execution with in-process boundaries, without a network or an
independent repetition. The evidence therefore characterizes the present cost but does not establish
variance, throughput, transport overhead, or suitability for a particular clinical or research
service. Practicality remains a deployment-specific question rather than a consequence inferred
from the affirmative feasibility result.


\section{Limitations}
\label{sec:limitations}

The strongest measurement is one complete execution of all $12$ released transformer blocks and
the genomic-signal task head for one held-out input at the full $103$-token prompt length. It
establishes numerical agreement for that classifier execution, including the final prediction, but
does not estimate encrypted task accuracy over the full test set. The complete run has one clean
sample and has not been independently repeated to quantify run-to-run variation.

Client and server are roles inside one program on one node. Boundary calls perform cryptographic
operations directly, without ciphertext serialization, network transport, or a key-custody service.
A physical client boundary crossing is therefore not a network message. A deployment experiment
must measure payload volume, dependency phases, bandwidth sensitivity, transport latency, and
production key handling rather than infer them from the in-process counter.

Encryption begins at embedded numeric vectors. Tokenization, token-index lookup in the embedding
table, and construction of the initial embedding occur before the encrypted boundary. Private
lookup therefore remains unresolved even though the classifier computation after embedding now
completes.

Model confidentiality remains operational rather than cryptographic. Because the data owner sees
full activations at every nonlinearity, the transcript divides the network into shallow segments
with known nonlinearities. An attacker can then recover each segment's weights by ordinary
regression, rather than by inverting the whole model at once. Server-side weight placement can
discourage casual copying but cannot
protect the model from a determined client; the deployment keeps the model on the server because
the server operates it.

CKKS decryption is approximate. The client never returns a decoded value to the server, only a
freshly encrypted result, but the server selects the ciphertext supplied to each declared boundary.
The semi-honest threat model rules out deviations from that schedule. The current re-encryption
path applies no noise flooding. The measured numerical headroom does not establish a secure
flooding parameter, which remains future work. The adversary exclusions are defined in
Section~\ref{sec:threat-model}; the results do not extend beyond them.


\section{Conclusion}

Client-assisted CKKS executes all $12$ released DNAGPT transformer blocks and the genomic-signal
task head for one held-out $103$-token input. Intermediate errors remain within
$1.54\times10^{-7}$, the encrypted head margin has $8.56\times10^{-9}$ relative error, and the
final label matches the independently validated plaintext reference. The complete execution peaks
at $9{,}839$~MiB of process GPU memory. Arithmetic feasibility therefore holds from encrypted
embedded vectors through the released head.

The measured cost is $6{,}683$~s for one example on a single accelerator. Server-side encrypted
linear algebra accounts for $85.5\%$ of encrypted evaluation and CPU-only client boundaries for
$14.5\%$. These measurements characterize present cost without establishing deployment
practicality: independent-run variation and network transport remain unmeasured, and private
token-index lookup remains outside the encrypted boundary.

The comparison with non-interactive CKKS identifies architecture, rather than numerical accuracy,
as the decisive constraint. Polynomial nonlinearities completed one real-weight block, but their
composition configuration exceeded the tested $80$~GB GPU-memory envelope. Exact nonlinearities and
fresh encryption at declared client boundaries bound the depth of each encrypted segment.
Separately, encoding public weight diagonals once and evicting their templates at stage boundaries
bounds host memory without changing the checked operation schedule.

\subsection*{Future work}

The immediate measurement priority is an independent repetition of the complete execution, followed
by a serialized client/server transport experiment. Systems work should cache fixed mask plaintexts
and evaluate distinct transforms across packed activation copies. A thread-safe cryptographic
context could expose independent client boundaries to concurrency. Protocol closure also requires a
validated noise-flooding budget for re-encryption and private token-index lookup.


\section*{Ethics Statement}

This study involved no human participants, specimens, or private clinical records. It used public
reference datasets with recorded provenance. The work makes no clinical-performance claim and
does not present the prototype as a deployable privacy guarantee beyond its stated threat model.

\section*{Data Availability Statement}

All evaluation datasets are public. Their original sources, versions, preprocessing, and any
recovery route used are recorded in the repository's data-provenance document. Sequence data,
model weights, and cryptographic keys are not redistributed.

\section*{Code Availability Statement}

The evaluation harnesses, encrypted-evaluation implementation, environment definitions, and the
evidence record for the paper are maintained at
\url{https://github.com/Georgakopoulos-Soares-lab/private_genomic_dnagpt}. No repository-level
reuse license has yet been assigned.

\section*{Conflict of Interest Statement}

The authors declare no competing financial or non-financial interests.

\section*{Acknowledgment}

This work has been supported by the National Institute of General Medical Sciences of the
National Institutes of Health [R35GM155468 to I.G.S.]; start-up funds awarded to I.G.S.; and the
Texas POC Award (2026) to I.G.S.

\FloatBarrier
\bibliography{refs}

\end{document}